\documentclass[12pt,draftclsnofoot,onecolumn,journal]{IEEEtran}
\ifCLASSINFOpdf
   \usepackage[pdftex]{graphicx}
   \DeclareGraphicsExtensions{.pdf,.jpeg,.png}
\else
\fi
\usepackage{amsmath,dsfont,bbm,epsfig,amssymb,amsfonts,amstext,verbatim}\usepackage{amsopn,cite,subfigure}
\usepackage{multirow,multicol,lipsum,xfrac}
\usepackage{amsthm}
\usepackage{mathtools,amsthm}
\usepackage{perpage}
\usepackage{balance}
\usepackage{url}
\usepackage{amsfonts}
\usepackage{epsfig}
\usepackage[font={small}]{caption}
\usepackage{graphicx}
\usepackage{psfrag}	
\usepackage{etoolbox}
\usepackage{algorithmicx}
\usepackage[Algorithm,ruled]{algorithm}
\usepackage{algpseudocode}
\usepackage{pifont}
\usepackage[utf8]{inputenc}
\usepackage[T1]{fontenc}  
\usepackage[nolist]{acronym}
\usepackage{paralist}
\usepackage{enumitem}
\usepackage{bbm}
\usepackage[process=auto]{pstool}
\usepackage{tikz,pgfplots}
\usetikzlibrary{shapes,arrows}
\usepackage{pdfpages}

\newcommand{\setR}{\mathbb{R}}

\newcommand{\setW}{\mathbbmss{W}}

\newcommand{\setO}{\mathbbmss{O}}
\newcommand{\setC}{\mathbb{C}}

\newcommand{\snr}{\mathrm{SNR}}

\newcommand{\diag}[1]{\mathrm{diag} \left( #1 \right) }
\newcommand{\esnr}{\mathrm{ESNR}}

\newcommand{\rmF}{\mathrm{F}}

\newcommand{\rmj}{\mathrm{j}}

\newcommand{\her}{\mathsf{H}}

\newcommand{\man}{\mathcal{N}}

\newcommand{\maC}{\mathcal{C}}

\newcommand{\mm}{\mathrm{m}}
\newcommand{\ee}{\mathrm{e}}

\newcommand{\bh}{{\mathbf{h}}}

\newcommand{\bx}{{\boldsymbol{x}}}

\newcommand{\set}[1]{\left\lbrace#1\right\rbrace}
\newcommand{\brc}[1]{\left( #1 \right)}
\newcommand{\dbc}[1]{\left[ #1 \right]}

\newcommand{\bg}{{\mathbf{g}}}
\newcommand{\bs}{{\boldsymbol{s}}}

\newcommand{\bw}{{\mathbf{w}}}
\newcommand{\bww}{\mathbf{w}}
\newcommand{\bff}{\mathbf{f}}

\newcommand{\loge}{\mathrm{ln}}

\newcommand{\be}{{\mathbf{e}}}

\newcommand{\trp}{\mathsf{T}}

\newcommand{\mW}{\mathbf{W}}

\newcommand{\mP}{\mathbf{P}}

\newcommand{\mI}{\mathbf{I}}
\newcommand{\mone}{\mathbf{1}}

\newcommand{\mG}{\mathbf{G}}

\newcommand{\mTheta}{\mathbf{\Theta}}
\newcommand{\mU}{\mathbf{U}}

\newcommand{\mV}{\mathbf{V}}

\newcommand{\mB}{\mathbf{B}}

\newcommand{\mH}{\mathbf{H}}
\newcommand{\mF}{\mathbf{F}}

\newcommand{\avg}{\mathrm{sum}}

\newcommand{\Ex}[2]{\mathbbm{E}_{#2}\left\lbrace #1 \right\rbrace}
\newcommand{\Var}[1]{\mathbbm{V}\left\lbrace #1 \right\rbrace}

\newcommand{\BB}{{\mathrm{BB}}}

\newcommand{\sinr}{{\mathrm{SINR}}}

\newcommand{\argmin}{\mathop{\mathrm{argmin}}}

\newcommand{\norm}[1]{\lVert #1 \rVert}

\newcommand{\abs}[1]{\left\lvert #1 \right\rvert}
\newcommand{\tr}[1]{\mathrm{tr} \left\lbrace #1 \right\rbrace }

\newtheoremstyle{mystyle}
  {}
  {}
  {\it}
  {}
  {\bfseries}
  {:}
  { }
  {}

\theoremstyle{mystyle}

\newtheorem{definition}{Definition}
\newtheorem{theorem}{Proposition}
\newtheorem{remark}{Remark}

\newtheorem*{protocol}{TAS Protocol}
\newtheorem*{scheme}{HADP Scheme}

\algnewcommand\algorithmicLet{\textbf{Let}}
\algnewcommand\Let{\item[\algorithmicLet]}
\algnewcommand\algorithmicSet{\textbf{Set}}
\algnewcommand\Set{\item[\algorithmicSet]}

\algnewcommand\algorithmicInitiate{\textbf{Initiate}}
\algnewcommand\Initiate{\item[\algorithmicInitiate]}
\algnewcommand\algorithmicStart{\textbf{Begin}}
\algnewcommand\Begin{\item[\algorithmicStart]}
\algnewcommand\algorithmicEnd{\textbf{End}}
\algnewcommand\End{\item[\algorithmicEnd]}

\algnewcommand\algorithmicOutP{\textbf{Output:}}
\algnewcommand\Out{\item[\algorithmicOutP]}

\algnewcommand\algorithmicInP{\textbf{Input:}}
\algnewcommand\In{\item[\algorithmicInP]}

\newcounter{bar}

\begin{document}
\title{Robustness of Low-Complexity Massive MIMO Architectures Against Passive Eavesdropping}

\author{
\IEEEauthorblockN{
Ali Bereyhi, \textit{Student Member, IEEE},
Saba Asaad, \textit{Student Member, IEEE},
Ralf R. M\"uller,  \textit{Senior Member, IEEE},
Rafael F. Schaefer, \textit{Senior Member, IEEE},
Georg Fischer,  \textit{Senior Member, IEEE}, and H.~Vincent~Poor, \textit{Fellow, IEEE}
\thanks{This work has been presented in parts at 2018 IEEE Global Communications Conference (GLOBECOM) in Abu~Dhabi \cite{bereyhi2018robustness}.}
\thanks{Saba Asaad, Ali Bereyhi and Ralf R. M\"uller are with the Institute for Digital Communications, Friedrich-Alexander Universit\"at Erlangen-N\"urnberg, Germany, \textit{\{saba.asaad, ali.bereyhi, ralf.r.mueller\}@fau.de}. Rafael F. Schaefer is with the Information Theory and Applications Chair, Technische Universit\"at Berlin, Germany, \textit{rafael.schaefer@tu-berlin.de}. 
Georg Fischer is with the Institute for Electronics Engineering, Friedrich-Alexander Universit\"at Erlangen-N\"urnberg, Germany, \textit{georg.fischer@fau.de}. 
H.~Vincent~Poor is with the Electrical Engineering Department, Princeton University, NJ 08544, \textit{poor@princeton.edu}.}
}
}
\IEEEoverridecommandlockouts

\maketitle

\begin{abstract}
Invoking large transmit antenna arrays, massive MIMO wiretap settings are capable of suppressing passive eavesdroppers via narrow beamforming towards legitimate terminals. This implies that secrecy is obtained almost \textit{for free} in these settings. We show that this property holds not only for fully digital MIMO architectures, but also in massive MIMO settings whose transmitters employ architectures with reduced complexity. The investigations consider two dominant approaches for complexity reduction, namely \textit{antenna selection} and  \textit{hybrid analog-digital precoding}. We show that using either approach, the information leakage normalized by the achievable sum-rate vanishes as the transmit array size grows large. For both approaches, the decaying speed is determined. The results demonstrate that, as the transmit array size grows large, the normalized leakages obtained by antenna selection and hybrid analog-digital precoding converge to zero \textit{double-logarithmically} and \textit{logarithmically}, respectively. These analytic derivations are confirmed for various benchmark architectures through numerical investigations. 
\end{abstract}
\begin{IEEEkeywords}
Massive MIMO systems, physical layer security, antenna selection, hybrid analog-digital precoding.
\end{IEEEkeywords}

\IEEEpeerreviewmaketitle

\section{Introduction}
Ensuring security and privacy of data during transmission is generally a core issue in wireless networks. This issue is more dominant in future wireless networks, i.e., 5G and beyond. In fact, due to the increasing demand of data rate, these networks are planned to have prominent features, such as high spectral efficiency, low latency and reduced cost. Consequently, they are expected to support diversified services, such as e-health, e-banking, and e-learning \cite{andrews2014will,nunna2015enabling,boccardi2014five,li20185g,xia20185g,david20186g}. These services require higher levels of privacy and security which imply additional secrecy challenges.

Traditional approaches based on cryptographic techniques at the higher layers utilize complicated key generation and distribution methods to provide reliability, security and data integrity \cite{walters2007wireless}. Nevertheless, staggering growth in the number of smart devices with powerful computational capabilities surmounts the conventional cryptographic approaches and a new paradigm under the generic term of \textit{physical layer security} has emerged as an complementary solution. The physical layer security approach, which relies on properties of the underlying wireless medium, exploits imperfections of the radio channel to provide a well-integrated secure platform for communication~\cite{yang2015safeguarding}. 

The seminal work of Wyner fundamentally laid an information-theoretic foundation of physical layer security by proving that in a point-to-point wiretap channel secure data transmission is achievable without secret key sharing, as long as the quality of the channel to the legitimate user is better than the one to the eavesdropper \cite{wyner1975wire}. Since then, physical layer security has been studied in a much broader scale, and the viewpoint of Wyner has been 
applied to more general scenarios \cite{liang2009information,shiu2011physical,poor2017wireless}. Some examples are the non-degraded versions of broadcast channels \cite{csiszar1978broadcast}, the Gaussian wiretap channels \cite{leung1978gaussian}, and relay channels \cite{oohama2001coding}. The analysis has been further extended to fading channels \cite{gopala2008secrecy,barros2006secrecy} and \ac{mimome} wiretap settings \cite{khisti2010secure,oggier2011secrecy}, in the more recent literature.

\subsection{Secrecy Robustness of Massive MIMOME Systems}
Recently, massive \ac{mimo} systems have received a great deal of attention \cite{marzetta2010noncooperative,hoydis2013massive,larsson2014massive}. Employing these systems in the initial version of 5G New Radio (5G NR) by 3GPP has further highlighted the importance of this technology for the next generations of wireless networks \cite{parkvall2017nr}. Beside unprecedented enhancement in spectral and energy efficiency, deploying a large number of antennas at the transmitter or receiver can provide promising secrecy opportunities. Specifically, a transmitter with a large antenna array is capable of narrowing the radiation beam pattern towards intended receivers, and hence attenuating the signal power in other diretions \cite{kapetanovic2015physical,schaefer2017physical}. In this case, a simple eavesdropper which is located outside the beam point receives almost no signal. This means that in this case a secure end-to-end transmission is achieved by utilizing the so-called \textit{favorable propagation property} which guarantees \textit{asymptotic orthogonality} of massive \ac{mimo} channels \cite{ngo2014aspects,wu2017favorable}.

The robustness of massive \ac{mimome} settings against \textit{passive} eavesdropping was initially demonstrated for \textit{not-too-dense} networks in \cite{kapetanovic2015physical} via numerical investigations, and later in \cite{schaefer2017physical} analytically. The impact of passive eavesdropping on the secrecy performance in \textit{dense networks} was investigated in \cite{wang2016secrecy}. Massive \ac{mimome} settings with \textit{active} malicious terminals were further studied in \cite{zhu2014secure,schaefer2017physical}. For these settings, secrecy performance was shown to be significantly degraded. The degradation can be partially compensated by means of artificial noise generation~\cite{zhu2014secure}.

Similar to other performance gains of massive \ac{mimo}, the robustness against passive eavesdropping is obtained at the expense of high implementational cost and complexity imposed by the large number of antennas in these systems. As fully~digital~implementation of massive \ac{mimo} systems does not seem to be tractable in the near future, it is important to answer the following question: \textit{Are massive \ac{mimome} settings with reduced complexity architectures also robust against passive eavesdropping?} The main objective of this work is to answer this question.

\subsection{Massive MIMO Systems with Reduced Complexity}
There are two major approaches to reduce the complexity of massive \ac{mimo} systems. The first approach is \ac{tas} in which downlink transmission is carried out through a subset of transmit antennas \cite{molisch2005capacity,li2014energy,benmimoune2015joint,asaad2017tas,asaad2018massive,bereyhi2017asymptotics,bereyhi2019glse}. Some recent studies have demonstrated that in addition to complexity reduction, \ac{tas} can provide performance enhancements in various other respects; see for example \cite{zhu2016secrecy,asaad2017optimal,asaad2018optimal,bereyhi2018stepwise}. The main alternative to \ac{tas} for complexity reduction is \ac{hadp}. In this approach, digital precoding is performed over a reduced number of \ac{rf} chains. The outputs of the \ac{rf} chains are then coupled via an analog network and transmitted over the transmit antenna array \cite{alkhateeb2013hybrid,el2014spatially,alkhateeb2014mimo,molisch2017hybrid}. Depending on the cost and complexity constraints, there exist various structures for digital~and~analog~beamforming, e.g., \cite{liang2014low,rusu2016low,gao2016energy,sedaghat2017novel,gade2018hybrid,gade2019outphasing}. The most extreme case is the \textit{single-\ac{rf} architecture} in which the transmit signal construction is completely performed in the \ac{rf} domain via the analog network \cite{kalis2008novel,sedaghat2014novel,sedaghat2016load}.

Despite the differences, \ac{tas} and \ac{hadp} can be abstractly presented via a unique \textit{hybrid structure}. In this structure, the information symbols are first mapped into a low-dimensional \ac{rf} signal \textit{digitally}. The \ac{rf} signal is then mapped into a signal of higher dimension via \textit{analog beamforming}. The \ac{tas} and \ac{hadp} are special instances of such a hybrid structure, where analog beamforming is realized in these examples via \textit{\ac{rf} switches}  and \textit{phase shifters}, respectively.

\subsection{Contributions and Organization}
The main objective of this study is to investigate the robustness of massive \ac{mimome} settings against passive eavesdropping. We address this objective through the following contributions:
\begin{itemize}
\item For a generic hybrid structure, we formulate mathematically the robustness against passive eavesdropping by introducing the concept of \textit{asymptotic secrecy for free}. To this end, we define the \textit{relative secrecy cost} in a wiretap setting. In settings which are robust against passive eavesdropping, this metric takes values close to zero.
\item Using the proposed formulation, we first investigate massive \ac{mimome} settings under \ac{tas}. We show that even by a \textit{fixed} number of active transmit antennas and linear digital precoding, the relative secrecy cost converges to zero proportional to $1/\log\log M$ where $M$ denotes the total number of transmit antennas. We confirm our analytic derivations by performing various numerical experiments.
\item We then extend the results to massive \ac{mimome} settings with \ac{hadp}. In this respect, we derive the relative secrecy cost as a function of system dimension in the large-system limit. The derivations show that in this case the cost tends to zero proportional to $1 / \log M$, even if the hybrid architecture employs \textit{only analog beamforming}. This analytic result is further supported via numerical investigations of some benchmark \ac{hadp} architectures.
\end{itemize}

The remaining parts of this manuscript are organized as follows: First, we formulate the problem and specify the system model in Section~\ref{sec:formulation}. The concept of secrecy for free is then introduced in Section~\ref{sec:SFF}. The robustness of massive \ac{mimome} settings against passive eavesdropping is then investigated under \ac{tas} and \ac{hadp} in Sections~\ref{sec:tas} and \ref{sec:hadp}, respectively. Final remarks and concluding points are further given in Section~\ref{sec:Conc}.

\subsection{Notations}
Throughout the paper, scalars, vectors and matrices~are~represented by~non-bold, bold lower case and bold upper case letters, respectively. The~set~of real numbers is denoted by $\setR$ and the complex plane is shown by $\setC$. $\mH^{\her}$, $\mH^{*}$ and $\mH^{\trp}$ indicate the Hermitian,~complex~conjugate and transpose of $\mH$, respectively. $\log\left(\cdot\right)$ and $\loge\brc\cdot$ represent the binary and natural logarithms, respectively. The expected value and variance of $x$ are denoted by $\Ex{x}{}$ and $\Var{x}{}$, respectively. The non-negative part of $x$ is shown by $\dbc{x}^+ \coloneqq \max\set{0, x}$. $\man\brc{\eta, \sigma^2}$ and $\mathcal{CN} \brc{\eta, \sigma^2}$ represent the real and complex~Gaussian~distributions with mean $\eta$ and variance $\sigma^2$, respectively. $\setO\brc\cdot$ is the \textit{big-O notation} describing the order of growth. For simplicity, $\set{1,\ldots,N}$ is shortened as $\dbc{N}$.

\section{Problem Formulation}
\label{sec:formulation}
Consider secure downlink transmission in a multiuser network with $K$ spatially distributed legitimate terminals and $J$ eavesdroppers. For sake of tractability, we assume that this network has a single \ac{bs} which is equipped with a transmit antenna array of size $M$ and $L$ \ac{rf} chains. The receiving terminals are further assumed to be single-antenna. 

\subsection{Channel Model}
\label{sec:sys}
The downlink channels from the \ac{bs} to the receiving terminals experiences a \textit{quasi-static fading} process. This means that the channel state remains constant during each coherence time interval which is longer than the transmission time. We focus on transmission in a single frequency sub-channel whose bandwidth is considerably narrower than the coherence bandwidth of the channels. This models either a single-carrier \textit{narrow-band} scenario or a particular sub-carrier of a multi-carrier system in a \textit{wide-band} setting. 

The system is assumed to operate in standard \ac{tdd} mode, as it is the case in \ac{mimo} settings with large antenna arrays. This means that the uplink and downlink channels are reciprocal. Following this reciprocity, the \ac{bs} estimates the \ac{csi} of the legitimate terminals by exploiting their \textit{pilot} sequences. These sequences are transmitted in the uplink training period at the beginning of each coherence time interval. We hence assume that the \ac{bs} knows the \ac{csi} of the legitimate users. The \ac{csi} of the malicious terminals are however not necessarily available at the \ac{bs}, as they are overhearing \textit{passively}.

Following the above illustrations, the uplink channel from the legitimate user $k$ to the \ac{bs} is modeled by $\left. \sqrt{\beta_k} \right. \bh_k$ for $k\in\dbc{K}$, where ${\beta_k}$ is capturing the path-loss and shadowing effects in the channel, and $\bh_k\in \setC^{M}$ models the fading process. Due to the reciprocity, entries of $\sqrt{\beta_k} \bh_k$ also denote the coefficients of the downlink channel from the \ac{bs} to legitimate terminal $k$. Similarly, the uplink channel from eavesdropper $j$ for $j\in\dbc{J}$ is given by $\left. \sqrt{\theta_j} \right. \bg_j$ where ${\theta_j}$ accounts for the path-loss and shadowing effect, and $\bg_j\in\setC^M$ captures the fading. For sake of brevity, we define the \textit{legitimate uplink channel matrix} to be $\mH \sqrt{\mB}$ and the \textit{overhearing uplink channel matrix} to be $\mG \sqrt{\mTheta}$, where $\mB \coloneqq \diag{\beta_1,\ldots,\beta_K}$ and $\mTheta \coloneqq \diag{\theta_1,\ldots,\theta_J}$ are diagonal matrices, and 
\begin{subequations}
\begin{align}
\mH &\coloneqq \dbc{\bh_1, \ldots, \bh_K},\\
\mG &\coloneqq \dbc{\bg_1, \ldots, \bg_J}.
\end{align}
\end{subequations}

For the fading process, we consider the standard Rayleigh model. This means that the entries of $\bh_k$ and $\bg_j$ for $k\in\dbc{K}$ and $j\in\dbc{J}$ are \ac{iid} complex Gaussian random variables with zero-mean and unit variance, i.e., $\bh_{k} , \bg_{j} \sim \mathcal{CN}\brc{\boldsymbol{0}, \mI_M}$. The investigations given in the next parts of this paper are straightforwardly extended to various other models of fading channels. Furthermore, we define the functions $\rmF\brc{\beta}$ and $\rmF\brc{\theta}$ to be the empirical cumulative distributions of $\beta_1,\ldots,\beta_K$ and $\theta_1,\ldots,\theta_J$, respectively. We hence use the notation $\Ex{f\brc{\beta}}{\rmF\brc{\beta}}$ and $\Ex{f\brc{\theta}}{\rmF\brc{\theta}}$ to denote respectively the empirical average of $f\brc{\beta_1},\ldots,f\brc{\beta_K}$ and $f\brc{\theta_1},\ldots,f\brc{\theta_J}$.

\subsection{Downlink Transmission via a Hybrid Structure}
The \ac{bs} intends to transmit a confidential message $u_k \in \dbc{ 2^{NR_k} }$ for $k\in\dbc{K}$ to legitimate terminal $k$ within $N$ channel uses, such that it is kept secret from the $J$ eavesdroppers. To this end, the \ac{bs} first encodes the secret message $u_k$ into the codeword 
$\dbc{s_k\brc{1}, \ldots, s_k\brc{N}}$ using encoder 
$f_{k,N}\brc{\cdot}:  \dbc{ 2^{NR_k} } \mapsto \setC^N$. %
It then maps the vector of \textit{encoded symbols} in time interval $n\in\dbc{N}$, i.e., %
$\bs\brc{n} = \dbc{s_1\brc{n}, \ldots, s_K\brc{n}}^\trp$,
into transmit signal $\bx\brc{n}\in\setC^M$ via a \textit{generic hybrid structure}: The \ac{bs} first constructs the \textit{base-band transmit signal} $\bx_{\BB} \brc{n}$ via a \textit{linear digital precoder}. 
Let $\bw_k\in\setC^L$ be the \textit{digital beamformer} of legitimate terminal $k$ and $P_k$ denote the power allocated to this terminal. Defining the \textit{digital beamforming matrix} $\mW = \dbc{ \bw_1 , \ldots,\bw_K}\in\setC^{L\times K}$ and the \textit{power allocation matrix} $\mP = \diag{P_1,\ldots,P_K} \in \setR^{K\times K}$, the base-band transmit signal is compactly shown as $\bx_{\BB} \brc{n}= \mW \sqrt{\mP} \left. \bs\brc{n} \right.$.

To address the constraint on the transmit power, $\bw_k$ and $\mP$ are restricted to satisfy $\norm{\bw_k}^2 = 1$ and $\tr{\mP} \leq P$ for some non-negative real $P$. Assuming that encoded symbols are distributed Gaussian with zero mean and unit variance, i.e. $s_k\brc{n}\sim \mathcal{CN}\brc{0,1}$ for $k\in\dbc{K}$, these restrictions yield to the following upper bound on the transmit power:
\begin{align}
\Ex{\norm{\bx_{\BB} \brc{n}}^2 }{} \leq P.
\end{align}
Using the \ac{rf} chains, the base-band transmit signal is upconverted to the carrier frequency. The outputs of these chains are then coupled via \textit{analog beamformers} $\bff_\ell\in\setC^M$ for $\ell \in \dbc{L}$. 
By defining the \textit{analog precoding matrix} $\mF$ as $\mF \coloneqq \dbc{\bff_1,\ldots,\bff_L}$, the signal transmitted by the \ac{bs} in the time interval $n$ is finally written as \cite{alkhateeb2014mimo}
\begin{align}
\bx \brc{n}= \mF \mW \sqrt{\mP} \left. \bs\brc{n} \right. .
\end{align}
We assume that the analog network is \textit{loss-less} meaning that the transmit power after analog precoding remains the same. Thus, the analog beamformers satisfy $\norm{\bff_\ell}^2=1$ for $\ell\in\dbc{L}$. 

Considering a standard Gaussian wiretap setting for downlink transmission, the signal received at legitimate terminal $k$ is written as
\begin{align}
y_k\brc{n} = \left. \sqrt{\beta_k} \right. \bh_k^\trp \bx\brc{n} + v_k\brc{n}
\end{align}
for $n\in\dbc{N}$, where $v_k\brc{n}\sim\mathcal{CN}\brc{0,\sigma^2}$ denotes the $n$-th sample of \ac{awgn} at legitimate terminal $k$. Similarly, the signal overheard by the $j$-th eavesdropper in the $n$-th transmission time interval is given by
\begin{align}
z_j\brc{n} = \left. \sqrt{\theta_j} \right. \bg_j^\trp \bx\brc{n} + b_j\brc{n}
\end{align}
where $b_j\brc{n}\sim\mathcal{CN}\brc{0,\rho^2}$ is the $n$-th sample of \ac{awgn} at eavesdropper $j$.

\subsection{Special Cases of the Hybrid Structure}
Depending on the choice of digital and analog beamformers, the given generic hybrid structure describes various low-complexity \ac{mimo} transmission schemes. In the sequel, we focus on two well-known techniques, namely \textit{\ac{tas}} and \textit{standard \ac{hadp}}. For each of these  techniques, the analog and digital beamforming matrices are of the following forms:

\subsubsection{TAS} Analog beamforming in \ac{tas} is implemented via a switching network. At beginning of each coherence time interval, a \textit{\ac{tas} protocol} selects $L$ transmit antennas to be active. The outputs of the \ac{rf} chains are then connected through the switching network to these antennas for the given coherence time interval. Let $i_1 < \ldots < i_L$ denote the indices of the selected transmit antennas. The analog beamformers in this case are sparse vectors with a single non-zero entry,~i.e., %
$\mathrm{f}_{\ell,m} = \mone\set{m = i_\ell}$, %
where $\mone\set{\cdot}$ denotes the indicator function. 

The digital beamformers are further calculated via standard linear precoding, e.g., \ac{mrt} or \ac{rzf}, over the \textit{reduced} channels between the legitimate terminals and the \ac{bs}, i.e., $\tilde{\bh}_k = \mF^\trp \bh_k$ for $k\in \dbc{K}$. For instance, in the case of \ac{mrt} precoding, digital beamformers are given by $\bw_k = {\tilde{\bh}_k^*} / { \norm{\tilde{\bh}_k} }$.

\subsubsection{HADP}
In \ac{hadp}, analog beamformers are realized via phase-shifters. The most conventional approach for implementation is to use a physical network of analog phase-shifters; see for example \cite{alkhateeb2013hybrid,liang2014low,el2014spatially,rusu2016low,gao2016energy, garcia2016hybrid,huang2018constant} and references therein. Some recent proposals for \ac{mmW} communications suggest to implement \ac{hadp} in \ac{mmW} spectrum via reflect- or transmit-arrays, e.g., \cite{jamali2019scalable,bereyhi2019papr}. Compared to conventional architectures, 
the latter structures have shown to have higher power efficiency and scalability \cite{jamali2019scalable}.

As analog beamformers are implemented only via phase-shifters, it is typically assumed~that transmit power is split uniformly across the transmit antennas. Hence, the $m$-th entry of beamformer $\bff_\ell$ for $m\in\dbc{M}$ is written in this case for some $\phi_{\ell,m}\in \setR$ as
\begin{align}
\mathrm{f}_{\ell,m} = \frac{1}{\sqrt{M}} \exp\set{\rmj \phi_{\ell,m} }. \label{eq:analog_HADP}
\end{align}

For digital beamforming, one can follow a similar approach as in \ac{tas} and employ a conventional linear digital precoder by considering the end-to-end channels between the \ac{rf} chains and the receiving terminals, i.e., $\mF^\trp \mH$. Alternatively, the digital beamformers can be designed directly via optimizing an objective measure of performance. Let $\mathcal{U}$ be a metric which quantifies the performance of the network. Such a metric is in general a function of the channel matrices, power allocation, as well as analog and digital beamformers. Hence, for a given setting and fixed analog beamformers, $\mathcal{U}$ is written as $\mathcal{U} = f\brc{ \mW,\mP }$. In this case, one can find the optimal digital beamformers with respect to $\mathcal{U}$ by optimizing $f\brc{ \mW,\mP }$ over the set of all possible choices. In practice, there are various metrics for performance characterization, e.g., \ac{lse}, energy efficiency and achievable throughput; see studies in \cite{khalid2014robust,mo2017hybrid} and references therein for further discussions. Some of these metrics are optimized via computationally tractable algorithms; however, for some others exact derivation is not tractably, and hence the beamformers are approximated by sub-optimal algorithms. In this paper, we assume that analog and digital beamformers are designed such that the secrecy performance is optimized. The metric with which the secrecy performance is quantified is given later~in~the sequel.

\subsection{Secrecy Performance}
To guarantee secure transmission, we need to ensure that there is no information leakage from the \ac{bs} to the eavesdroppers. Considering the \textit{worst-case} scenario in which the eavesdroppers are cooperating, achievable secrecy rates are defined as follows:

\begin{definition}[Achievable secrecy rate]
The secrecy rate tuple $\brc{R_1,\ldots,R_K}$ is said to be achievable, if there exists a sequence of encoders $f_{k,N}:  \dbc{ 2^{NR_k} } \mapsto \setC^N $ and decoders $\phi_{k,N}: \setC^N\mapsto  \dbc{ 2^{NR_k} }$, indexed by $N$, for $k\in\dbc{K}$ such that $\dbc{s_k\brc{1}, \ldots, s_k\brc{N}} = f_{k,N} \brc{u_k}$, and
\begin{subequations}
\begin{align}
&\lim_{N\uparrow \infty} \left. \max_{k\in\dbc{K}} \right. \Pr\set{\phi_{k,N} \brc{ y_k\brc{1}, \ldots, y_k\brc{N} } \neq u_k }  = 0 \label{eq:Const1}\\
&\lim_{N\uparrow \infty} \left. \frac{1}{N} \right. \mathrm{I} \brc{
\mathcal{S}\brc{u_1,\ldots,u_K};z_1\brc{1},\ldots,z_1\brc{N},z_J\brc{1},\ldots,z_J\brc{N}
}  = 0 \label{eq:Const2}
\end{align}
for all $\mathcal{S}\brc{u_1,\ldots,u_K} \subseteq \set{u_1,\ldots,u_K}$.
\end{subequations}
\end{definition}
The constraint in \eqref{eq:Const1} guarantees \textit{reliability} of downlink transmissions, while \eqref{eq:Const2} enforces the overheard signals to contain \textit{no information leakage}.

For legitimate terminal $k$, an achievable secrecy rate is given by \cite{oggier2011secrecy,ekrem2011secrecy,geraci2012secrecy}
\begin{align}
R_k \brc{M} = \dbc{\log \left( \frac{ 1+ \sinr_k\brc{M}}{ 1+ \esnr_k\brc{M} }  \right)}^+,
\label{eq:R_k}
\end{align}
where $\sinr_k\brc{M}$ and $\esnr_k\brc{M}$ are defined as follows
\begin{subequations}
\begin{align}
\sinr_k\brc{M} &= \frac{ P_k \beta_k \left. \abs{\bh^\trp_k \mF \bww_k}^2 \right.  }{ \sigma^2 + \beta_k \sum\limits_{i=1, i\neq k}^K P_i \left. \abs{\bh^\trp_k \mF \bww_i}^2 \right. },\\
\esnr_k\brc{M} &= \dfrac{P_k }{\rho^2} \left. \norm{\sqrt{\mTheta} \left. \mG^\trp  \mF \bww_k}^2 \right.\right. .
\end{align}
\end{subequations}
The achievability of the given secrecy rates requires the \ac{csi} of the eavesdroppers to be \textit{available} at the \ac{bs}. This assumption does not necessarily hold in the network. Hence, the secrecy rates in \eqref{eq:R_k} describes the secrecy performance of the system in a \textit{superior} condition in which a genie informs the \ac{bs} about the \ac{csi} of overhearing channels. It is later shown that the given secrecy performance is achievable without knowing the eavesdroppers' \ac{csi}. 

In the absence of malicious terminals, i.e., when $J=0$, the achievable secrecy rate reduces to
\begin{align}
R^{\rm m}_k \brc{M} = \log \left( 1+ \sinr_k\brc{M}  \right)
\label{eq:Rm_k}
\end{align}
which denotes the rate to user $k$ achieved by linear precoding in the standard downlink setting without eavesdropper. Unlike the secrecy rate, $R^{\rm m}_k \brc{M}$ is achieved via standard channel coding which does not require to know the \ac{csi} of overhearing channels. Using this approach for downlink transmission in a network with malicious terminals, the eavesdroppers receive some \textit{information leakage}. In \eqref{eq:R_k}-\eqref{eq:Rm_k}, the number of transmit antennas, i.e., $M$, is denoted as an argument to explicitly indicate the dependency of these expressions on the array size.

To quantify the secrecy performance in this network, we utilize the achievable rate in \eqref{eq:R_k} and define the following performance metric:

\begin{definition}[Weighted secrecy sum-rate]
The weighted achievable secrecy sum-rate is 
\begin{align}
R_{\avg} \brc{M} = \sum_{k=1}^K q_k \left. R_k\brc{M} \right.
\end{align}
for some weighting coefficients $q_1, \ldots, q_K$. 
\end{definition}

The weighting coefficient $q_k$ is proportional to the \ac{qos} desired for the particular legitimate terminal $k$. For instance, when all the users are supposed to meet the same \ac{qos}, the weights are constant, i.e., $q_k = q_\ell$ for $k\neq \ell\in \dbc{K}$. The achievable weighted downlink sum-rate in this case reads
\begin{align}
R^\mm_{\avg} \brc{M} = \sum_{k=1}^K q_k \left. R^\mm_k\brc{M} \right. .
\end{align}

To investigate the performance of the system in the large-system limit, we take the asymptotic limit of $M\uparrow\infty$. This means that we consider a sequence of \ac{mimome} settings, indexed by $M$, which are consistent with the model described in this section and assume that $M$ can grow unboundedly large. The numbers of legitimate terminals and eavesdroppers are set fixed. 

\section{Secrecy For Free}
\label{sec:SFF}
In \ac{mimome} settings with large transmit arrays, the \ac{bs} is able to suppress the signals received by the eavesdroppers via narrow beamforming towards legitimate terminals. Hence, when $M$ grows large, $R_k\brc{M}$ converges to $R^{\rm m}_k\brc{M}$ even in the presence of eavesdroppers which implies that secrecy is obtained at no significant cost. In other words, the information leakage to the eavesdroppers converges to zero even by employing a standard downlink transmission scheme which ignores the malicious terminals. We refer to this property as \textit{secrecy for free} and formulate it formally in the sequel. 

\subsection{Asymptotic Secrecy For Free}
Physical layer secrecy is obtained at the expense of reduction in the data rate. This can be seen by comparing the achievable rates in \eqref{eq:R_k} and \eqref{eq:Rm_k}. The comparison further indicates that the cost depends on the quality of the channels by which the eavesdroppers overhear the transmit signal. We quantify this cost by defining the \textit{relative secrecy cost} as follows:
\begin{definition}[Relative secrecy cost]
\label{def:rel_cost}
Let $R_\avg\brc{M}$ and $R^\mm_\avg\brc{M}$ denote the achievable weighted secrecy sum-rate in the presence of eavesdroppers in the network, i.e., $J\neq 0$, and the achievable weighted sum-rate in the absence of eavesdroppers, i.e., $J=0$, respectively. The relative secrecy cost in the network is defined as
\begin{align}
\maC \brc{M} \coloneqq 1-\dfrac{ R_\avg\brc{M} }{R^\mm_\avg\brc{M}},
\end{align}
when $R^\mm_\avg\brc{M} \neq 0$. The relative secrecy cost is defined to be $\maC \brc{M} = 0$, if $R^\mm_\avg \brc{M} =0$.
\end{definition}

$\maC\brc{M}$ quantifies the reduction in the achievable rate imposed intentionally by the \ac{bs} to secure the communication. From Definition~\ref{def:rel_cost}, it is straightforwardly inferred that %
$0 \leq \maC\brc{M} \leq 1$, %
where the lower and upper bounds on $\maC\brc{M}$ are met when $R_\avg \brc{M} = R_\avg^\mm \brc{M} $ and $R^\mm_\avg\brc{M}\neq R_\avg\brc{M}=0$, respectively. Following the intuitive definition, secrecy for free refers to scenarios in which $\maC\brc{M}$ tends to the lower bound when $M$ grows large. We hence define the concept of \textit{asymptotic secrecy for free} formally as follows:

\begin{definition}[Asymptotic secrecy for free]
\label{def:sec_free}
Consider a sequence of problems indexed by $M$.~Let transmit power $P$ be bounded from above. The secrecy is achieved asymptotically for free, if
\begin{align}
\lim_{M\uparrow \infty} \maC\brc{M} = 0.
\end{align}
\end{definition}

The above definition implies that, when secrecy is achieved asymptotically for free in the network, then we have $R_k \brc{M} \approx R_k^\mm \brc{M}$ for large transmit antenna arrays. This means that 
the \ac{bs} blinds the eavesdroppers without knowing the \ac{csi} of their channels. As a result, unlike the non-asymptotic regime, in the large-system limit, the \ac{bs} does not require the \ac{csi} of overhearing channels to provide physical layer security.

From practical points of view, the scale of \textit{massiveness} by which secrecy is almost for free in the network is not directly derived from the concept of secrecy for free. In fact, to achieve secrecy asymptotically for free, the relative secrecy cost only needs to drop monotonically with $M$; see Definition~\ref{def:sec_free}. Nevertheless, it is the speed of this drop which specifies the practical \textit{scale} of required system dimension. To address this concern, we further define the \textit{asymptotic decay of secrecy cost} as follows:

\begin{definition}[Asymptotic decay of secrecy cost]
Let $\maC\brc{M}$ denote the relative secrecy cost for a sequence of problems indexed by $M$ in which secrecy is achieved asymptotically for free. $f\brc{M}$ is the asymptotic decay of secrecy cost if  $\maC\brc{M} = \setO\brc{1/f\brc{M}}$, i.e., 
\begin{align}
\lim_{M\uparrow \infty} f\brc{M} \maC\brc{M} = C_0
\end{align}
for some bounded constant $C_0\in\setR$.
\end{definition}

The asymptotic decay of secrecy cost provides an approximate metric on the number of transmit antennas by which the system is considered \textit{massive enough} to achieve secrecy almost for free. In practice, we are interested in \ac{mimo} transmission schemes whose relative secrecy costs decay~fast.

In the remaining parts of this manuscript, we investigate the concept of secrecy for free in massive \ac{mimome} settings with reduced complexity, analytically. To this end, we consider the major low-complexity \ac{mimo} architectures, namely \ac{tas} and \ac{hadp}. For each of these architectures, we depict that secrecy is achieved asymptotically for free via a wide class of analog and digital beamforming techniques. The derivations show that using \ac{hadp} a higher asymptotic decay of secrecy cost is achievable with the same number of \ac{rf} chains. 

\section{Secrecy For Free under TAS}
\label{sec:tas}
We start the analyses by considering \ac{tas}. It is shown that even when only a subset of transmit antennas is set active, the relative secrecy cost converges to zero as the transmit array size grows large. For this architecture, the asymptotic decay of secrecy cost is at least $\log\log M$ which is rather slow. In the sequel, we first state the main result. The derivation of the result is then given in details and confirmed by some numerical simulations.

\subsection{Main Results}
The following proposition gives \textit{sufficient} conditions under which secrecy is achieved asymptotically for free via \ac{tas}. It further specifies a \textit{lower bound} on the asymptotic decay of secrecy~cost.

\begin{theorem}[Secrecy for free under TAS]
\label{theo:1}
Assume that the \ac{bs} uses only the \ac{csi} of legitimate channels for \ac{tas} and digital beamforming. Let analog beamformers be set via a \ac{tas} protocol which selects $L \geq K$ transmit antennas once per coherence time interval. Then, there exist linear digital beamformers which achieve secrecy asymptotically for free. The asymptotic decay of secrecy cost for this hybrid structure is at least $\log\log M$.
\end{theorem}

Proposition~\ref{theo:1} guarantees the existence of a \textit{linear} digital beamforming scheme and a \ac{tas} protocol which is updated at the \textit{Doppler rate}. The asymptotic decay of secrecy cost obtained by this \ac{tas} transmission scheme is at least $\log\log M$.  Throughout the derivations, we design a digital beamforming scheme and a \ac{tas} protocol which secrecy is achieved asymptotically for free, and $\maC\brc{M} = \setO\brc{1/\log\log M}$. Nevertheless, the result is not restricted to the proposed schemes and is valid for superior approaches, as well. 

\begin{remark}
It is worth to indicate that Proposition~\ref{theo:1} provides some \textit{sufficient} conditions and does not discuss any \textit{necessary} condition. This means that secrecy for free could be eventually achievable with smaller numbers of active transmit antennas. 
\end{remark}


\subsection{Derivations}
The proof of Proposition~\ref{theo:1} follows two major steps: First, we design \textit{linear digital beamformers}, as well as a \textit{\ac{tas} protocol} whose update rate is once per coherence time interval. Then, we show that for the proposed beamforming technique and \ac{tas} protocol, the relative secrecy cost converges to zero proportional to $1/\log \log M$. Throughout the derivations, we consider the case with exactly $L=K$ \ac{rf} chains. The result then immediately extends the proof to cases with $L \geq K$, since in settings with $L > K$ \ac{rf} chains, one can still set only $K$ antennas active.

\begin{protocol}
At the beginning of each coherence time interval, the \ac{bs} selects exactly $L=K$ antennas by pursuing the following steps:
\begin{enumerate}
\item For $k\in \dbc{K}$, the \ac{bs} sorts the channel gains from the transmit antennas to~legitimate~terminal $k$ in a decreasing order. This means that it finds indices $i_{k,1},\ldots,i_{k,M}\in\dbc{M}$, such that
\begin{align}
\abs{h_{k,i_{k,1}}}^2 \geq \ldots \geq \abs{h_{k,i_{k,M}}}^2
\end{align}
\item It then sets antennas $\ell_k$ for $k\in\dbc{K}$ active, where $\ell_1 = i_{1,1}$ and %
$\ell_k = i_{ k ,t_k }$ %
for $2\leq k \leq K$. Here, the index $t_k \in \dbc{M} $ is given by
\begin{align}
t_k \coloneqq \argmin_{a\in \setW_k} i_{k,a}
\end{align}
where
\begin{align}
\setW_k \coloneqq  \set{ w \in \dbc{M}: i_{k,w} \neq \ell_{k-b} \ \text{ for } \ b\in\dbc{k-1} }.
\end{align}
\end{enumerate}
\end{protocol}

In a nutshell, the \ac{tas} protocol selects $L=K$ antennas indexed by $\ell_k$ for $k\in\dbc{K}$ where $\ell_k$ denotes the antenna whose channel gain to the legitimate terminal $k$ is strongest. If the $Q$ strongest antennas are already taken by other users, $\ell_k$ is set to be the $Q+1$-th strongest antenna. Noting that the antennas are selected based on the \ac{csi} of the legitimate terminals, the protocol is updated once in a coherence time interval. Without loss of generality, we assume $\ell_1 < \ldots < \ell_K$.

Under this protocol, we have $L=K$ analog beamformers $\bff_1, \ldots, \bff_K$, where $\bff_k = \be^M_{\ell_k}$ with $\be^M_{\ell}\in \setR^M$ denoting the $\ell$-th principle basis vector\footnote{This means that the $\ell$-th entry of $\be^M_{\ell}$ is one, and the remaining entries are zero.}~of $\setR^M$. For digital beamforming, we use an \ac{mrt}-based scheme in which the $k$-th digital beamformer contains only the matched filter corresponding to the channel between antenna element $\ell_k$ and the $k$-th legitimate terminal, i.e.,
\begin{align*}
\bww_k = \frac{h_{k,\ell_k}^*}{\abs{h_{k,\ell_k}}}  \be_k^K.
\end{align*}

We now show that the relative secrecy cost for the given protocol and beamformers converges to zero asymptotically. To keep the derivations tractable, we assume that $\setW_k = \dbc{M}$ for all $k\in\dbc{K}$ meaning that the strongest antennas corresponding to different terminals are distinct. Extension to cases with overlapping strongest antennas is trivial. We start the derivations with determining the \textit{asymptotic expansions} of $\sinr_k\brc{M}$ and $\esnr_k\brc{M}$. Noting that
\begin{align}
\abs{\bh_k^\trp \mF \bw_i}^2 &= \abs{
\frac{h_{i,\ell_i}^*}{\abs{h_{i,\ell_i}}}
\bh_k^\trp \mF   \be_i^L }^2 = \abs{ \bh_k^\trp \bff_i }^2 =  \abs{ h_{k,\ell_i} }^2,
\end{align}
$\sinr_k\brc{M}$ reads
\begin{align}
\sinr_k \brc{M} &= \dfrac{ P_k \beta_k \left. \abs{ h_{k,\ell_k} }^2 \right.  }{ \sigma^2 + \beta_k I_k }
\label{eq:sinr_full}
\end{align}
where
\begin{align}
I_k \coloneqq \sum\limits_{i=1, i\neq k}^K P_i \left. \abs{ h_{k,\ell_i} }^2 \right. .
\end{align}
Similarly, for $\esnr_k \brc{M}$
\begin{align}
\left. \esnr_k \brc{M} \right. &= \dfrac{P_k }{\rho^2} \left. \norm{\sqrt{\mTheta} \left. \mG^\trp  \mF \bww_k}^2 \right.\right. = \dfrac{P_k E_k }{\rho^2}
\end{align}
where we define
\begin{align}
E_k \coloneqq \sum_{j=1}^J \theta_j \abs{ g_{j,\ell_k} }^2.
\end{align}

To pursue the derivations, we consider the following statements:
\begin{itemize}
\item[(a)] For $k\in \dbc{K}$, random variable $\abs{ h_{k,\ell_k} }^2$ reads
\begin{align}
\abs{ h_{k,\ell_k} }^2 = \max_{m\in\dbc{M}} \left.  \abs{ h_{k,m} }^2 \right.
\end{align}
which is an order statistics of $\bh_k$, often called as the \textit{extreme value} of $\bh_k$ \cite{arnold1992first,david2004order}. Noting that entries of $\bh_k$ are \ac{iid} Gaussian random variables, the \textit{Fisher-Tippett theorem} \cite{de2007extreme,gasull2015maxima} indicates that as $M$ grows large, the random variable $\Gamma_k$, defined as
\begin{align}
\Gamma_k \coloneqq \abs{ h_{k,\ell_k} }^2 - \loge M
\end{align}
converges in distribution to a Gumbel distribution with zero location and unit scale, i.e.,
\begin{align}
\Pr\brc{ \Gamma_k \leq \gamma } = \exp\set{-\exp\set{-\gamma}}.
\end{align}
This means that for $k\in \dbc{K}$ 
\begin{align}
\dfrac{\abs{ h_{k,\ell_k} }^2}{\loge M} \longrightarrow 1
\end{align}
where $\longrightarrow$ indicates convergence in \textit{mean square}.
\item[(b)] Since $\bh_i$ for $i \neq k$ are independent of $\bh_k$, entries $h_{k, \ell_i}$ are \textit{randomly} selected entries, and hence are Gaussian distributed with zero mean and unit variance. As a result, one can write
\begin{subequations}
\begin{align}
\Ex{I_k}{} &=  \sum\limits_{i=1, i\neq k}^K P_i  = P - P_k\\
\Ex{I_k^2}{} &=  \sum\limits_{i=1, i\neq k}^K P_i^2 + \sum\limits_{i=1, i\neq k}^K \sum\limits_{\ell=1, \ell\neq k}^K P_i P_\ell  = \sum\limits_{i=1, i\neq k}^K P_i^2 + \brc{P - P_k}^2.
\end{align}
\end{subequations}
Note that 
\begin{align}
\sum\limits_{i=1, i\neq k}^K P_i^2 \leq \sum\limits_{i=1}^K P_i^2 \stackrel{\dagger}{\leq} P^2,
\end{align}
where $\dagger$ follows from the fact that $P_i \geq 0$ for $i\in\dbc{K}$. This means that for any $M$, $I_k$ is a random variable with \textit{bounded} mean $\Ex{I_k}{}$ and \textit{bounded} variance
\begin{align}
\Var{I_k} =  \sum\limits_{i=1, i\neq k}^K P_i^2.
\end{align}
\item[(c)] With the same lines of justification as in the previous statement, it is concluded that $g_{j,\ell_k}$ are distributed Gaussian with zero mean and unit variance. Thus, we have
\begin{subequations}
\begin{align}
\Ex{ E_k }{} &= \sum_{j=1}^J \theta_j  = \left. J \right. \Ex{\theta}{\rmF\brc{\theta}}\\
\Ex{ E_k^2 }{} &= \sum_{j=1}^J \theta_j^2 + \brc{  \sum_{j=1}^J \theta_j}^2  = J \Ex{\theta^2}{\rmF\brc{\theta}} + \brc{J \Ex{\theta}{\rmF\brc{\theta}} }^2.
\end{align}
\end{subequations}
This concludes that for any $M$, $E_k$ is a random variable with \textit{bounded} mean $\Ex{E_k}{}$ and \textit{bounded} variance %
$\Var{E_k} =  J \Ex{\theta^2}{\rmF\brc{\theta}}$.
\end{itemize}

Using the expression derived for $\sinr_k\brc{M}$, we can write
\begin{align}
R_k^\mm \brc{M} = \log \brc{1 + \dfrac{ P_k \beta_k \left. \abs{ h_{k,\ell_k} }^2 \right.  }{ \sigma^2 + \beta_k I_k } } . 
\end{align}
Statement (a) indicates that as $M$ grows large, we have \cite{boyd1999devil}
\begin{subequations}
\begin{align}
R_k^\mm \brc{M} \longrightarrow &\log \brc{1 + \dfrac{ P_k \beta_k \left. \loge M \right.  }{ \sigma^2 + \beta_k I_k } } \\
= &\log \brc{  \loge M } + \log \brc{  P_k \beta_k} - \log \brc{ \sigma^2 + \beta_k I_k }  + \log \brc{ 1+ \dfrac{ \sigma^2 + \beta_k I_k }{ P_k \beta_k \left. \loge M \right.  } } \\
= &\log \brc{  \loge M } + \epsilon_k\brc{M},
\end{align}
\end{subequations}
where from Statement (b), it is concluded that $\epsilon_k\brc{M} = \setO\brc{1}$. 

For the achievable secrecy rate, we can further write
\begin{align}
R_k \brc{M} = \dbc{
\log \brc{1 + \dfrac{ P_k \beta_k \left. \abs{ h_{k,\ell_k} }^2 \right.  }{ \sigma^2 + \beta_k I_k } } -
\log \brc{1 + \dfrac{P_k E_k }{\rho^2} }
}^+.
\end{align}
Following Statements (a)-(c), we can conclude that for any realization of the channels, there exists $M_0$, such that for $M \geq M_0$
\begin{align}
\log \brc{1 + \dfrac{ P_k \beta_k \left. \abs{ h_{k,\ell_k} }^2 \right.  }{ \sigma^2 + \beta_k I_k } } \geq
\log \brc{1 + \dfrac{P_k E_k }{\rho^2} }
\end{align}
with probability one. Thus, in the large-system limit, we have
\begin{subequations}
\begin{align}
R_k \brc{M} \longrightarrow &\log \brc{1 + \dfrac{ P_k \beta_k \loge M  }{ \sigma^2 + \beta_k I_k } } -
\log \brc{1 + \dfrac{P_k E_k }{\rho^2} }\\
= &\log \brc{\loge M } + \hat{\epsilon}_k \brc{M}
\end{align}
\end{subequations}
with $\hat{\epsilon}_k \brc{M} = \setO\brc{1}$. As a result, the relative secrecy cost reads
\begin{align}
\label{eq:Asymp_TAS}
\maC \brc{M} \longrightarrow 
\dfrac{ 
\displaystyle \sum_{k=1}^K q_k \epsilon_k \brc{M} - \sum_{k=1}^K q_k \hat{\epsilon}_k \brc{M} 
}{
\displaystyle \log\brc{\loge M} + \sum_{k=1}^K q_k \epsilon_k \brc{M}
}.
\end{align}

From \eqref{eq:Asymp_TAS}, it concluded $\maC \brc{M} = \setO\brc{1/\log \log M}$. This indicates that secrecy is achieved asymptotically for free with the asymptotic decay of secrecy cost $\log\log M$ using only $K$ active antennas. The result ensures that the property is achievable with a larger number of \ac{rf} chains, i.e., for any $L\geq K$, with at least of the same decay speed. It is worth to indicate that $L\geq K$ is only a \textit{sufficient} condition. From the derivations, it is  observed that the proof is further extendable to scenarios with fewer \ac{rf} chains.

\begin{remark}
With some straightforward modifications, the proof extends to \textit{bi-unitarily invariant channel matrices}\footnote{The random matrix $\mH\in\setC^{M\times K}$ is \textit{bi-unitarily invariant}, if for any pair of independent unitary matrices $\mU\in\setC^{M\times M}$ and $\mV\in\setC^{K\times K}$, the entries of $\mH$ and $\mU\mH\mV^{\her}$ have the same distribution\cite{tulino2004random}.} which cover a wide class of fading models.
\end{remark}

\subsection{Numerical Investigations}
\label{eq:NumInv_TAS}
We now perform some numerical experiments to confirm the analytic results. To this end, we consider two distinct scenarios: In the first scenario, the network contains $K=16$ legitimate terminals and $J=2$ eavesdroppers uniformly distributed in a cell. As the number of malicious terminals is considerably smaller than the number of legitimate users, we refer to this scenario as the \textit{sparsely overheard} network. The second scenario considers a network with $K=16$ users and $J=16$ eavesdroppers. This network is referred to as the \textit{densely overheard} network. In both the networks, we assume that the \ac{bs} selects $L=K$ transmit antennas. 

Throughout the investigations, the \ac{snr}s at legitimate and malicious receiving terminals are defined as $\snr^\mm_k \coloneqq \beta_k P/\sigma^2$ for $k\in\dbc{K}$ and $\snr^\ee_j \coloneqq \theta_j P/\rho^2$ for $j\in\dbc{J}$, respectively. In simulations, we set $P=\sigma^2=\rho^2=1$. It is further assumed that the impact of path-loss and shadowing at legitimate terminals is compensated at the transmit side; hence, we set $\beta_k = 1$ for $k\in\dbc{K}$. For the eavesdroppers, we assume $\theta_i=0.1$ for $j\in\dbc{J}$. As a result, $\log \snr^\mm_k = 0$ dB and $\log \snr^\ee_j = -10$ dB for $k\in\dbc{K}$ and $j\in\dbc{J}$.

For each of the scenarios, we provide two sets of numerical results:
\begin{itemize}
\item In the first set, simulation results for the basic \ac{tas} transmission scheme, which is described through the derivations, is given. We refer to this scheme as \textit{\ac{tas} Scheme A}.
\item The simulations are then performed for a \textit{benchmark} \ac{tas} scheme given in \cite{bereyhi2018stepwise}. In this scheme, the active transmit antennas are selected via a \textit{step-wise} approach such that the weighted sum-rate is maximally increased in each step. For digital precoding over the selected antennas, we use the \ac{mrt} scheme. This means that $\bw_k = \tilde{\bh}_k^*/\norm{\tilde{\bh}_k}$, where $\tilde{\bh}_k$ denotes the $k$-th column of $\mF^\trp \mH$. We refer to this scheme as \textit{\ac{tas} Scheme B}. More details of this scheme can be found in \cite{bereyhi2018stepwise}.
\end{itemize}
We further assume uniform power allocation in the simulations for both the schemes, and set $q_1 = \ldots =q_K=1$ while calculating a weighted sum-rate.

\begin{figure}
\centering
\input{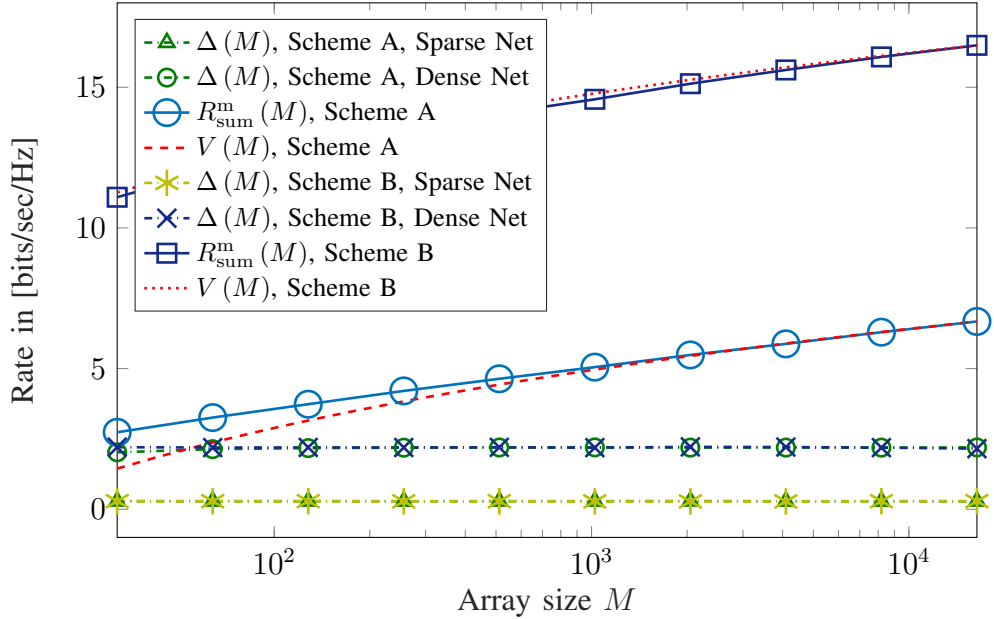}
\caption{Achievable sum-rate and information leakage in for both the \ac{tas} schemes.}
\label{fig:1}
\end{figure}

Fig.~\ref{fig:1} sketches the achievable sum-rate, as well as the \textit{information leakage} to the eavesdroppers, against the transmit array size for both the \ac{tas} schemes. Here, the information leakage is defined as %
$\Delta \brc{M} \coloneqq R_\avg^\mm \brc{M} - R_\avg \brc{M}$.
The achievable sum-rate is the same for the two networks, as the numbers of legitimate terminals are the same in both networks.

As Fig.~\ref{fig:1} shows, the leakage in both the schemes remains almost constant for all choices of $M$, i.e., $\Delta \brc{M} = \delta$ for some small $\delta$. This indicates that $R_\avg^\mm \brc{M} \approx R_\avg \brc{M}$, or in other words, secrecy is achieved almost for free in this case. In the densely overheard network, the amount of leakage is rather higher. This is due to the high density of the malicious terminals in the network. As a result, $R_\avg^\mm \brc{M} \approx R_\avg \brc{M}$ is an accurate approximation for larger $M$.

\begin{figure}
\centering
\input{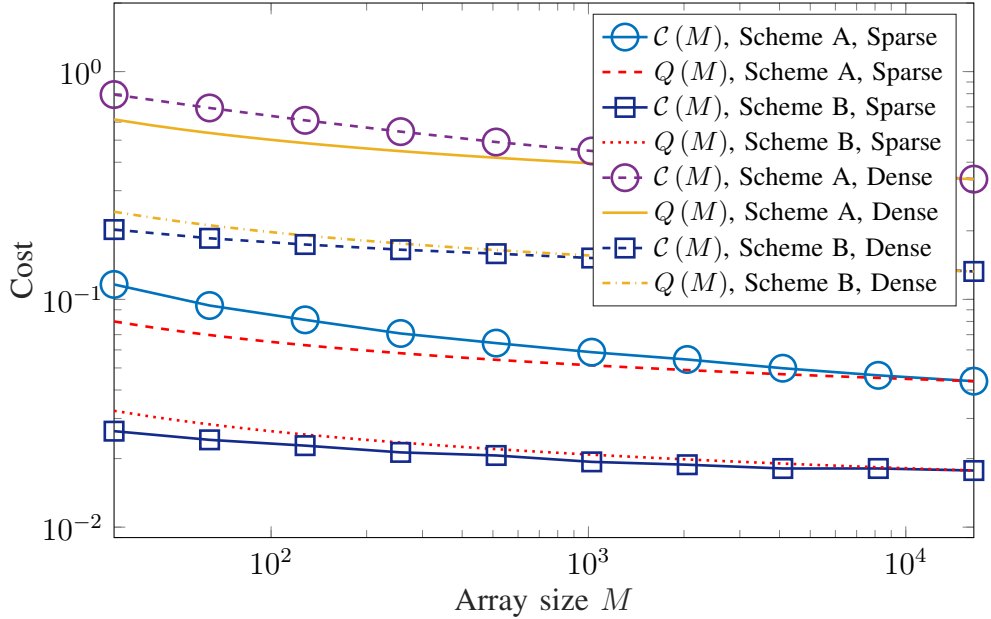}
\caption{Relative secrecy cost for both the \ac{tas} schemes.}
\label{fig:2}
\end{figure}

Unlike the leakage, the achievable sum-rate grows with $M$. This observation confirms the prior analysis. It is further seen that the achievable sum-rate in \ac{tas} Scheme B is shifted by approximately $9$ dB on the vertical axis compared to \ac{tas} Scheme A. This is as the benchmark scheme performs superior to the basic scheme considered for derivations. It is however observable from the figure that the order of growth is almost the same for the two schemes. Since the information leakage is constant in $M$, the asymptotic decay of secrecy cost indicates that the achievable downlink rate to each legitimate user grows in this case proportional to $\log\log M$. One can hence write in this case
\begin{align}
R_\avg^\mm \brc{M} \approx V \brc{M} = R_0 + T K \log \log M \label{eq:FM}
\end{align}
for some constants $T$ and $R_0$. For the given curves in Fig.~\ref{fig:1}, we find the values of $T$ and $R_0$ by curve fitting. The resulting curves are plotted with dashed and dotted lines in the figure. As it is seen, the curves track the numerical simulations asymptotically for \textit{both} the~schemes.~This means that the lower bound on the asymptotic decay of secrecy cost is also \textit{tight} for other \ac{tas} schemes.

The relative secrecy cost for the given schemes are further plotted in Fig.~\ref{fig:2} versus the array size. Considering the asymptotic decay of secrecy cost, we can write in the large-system limit
\begin{align}
\maC \brc{M} \approx Q \brc{M} = \frac{\epsilon_0}{\log \log M}
\end{align}
for some constant $\epsilon_0$. $Q \brc{M}$ is further plotted in both the figures for both the schemes. The value of $\epsilon_0$ for each curve is found such that $\maC \brc{M_0} = Q \brc{M_0}$ for $M_0 = 2^{14}$. As it depicts, the fitted curves track the simulation results.


%

From Fig.~\ref{fig:2}, it is observed that for achieving a desired relative secrecy cost, either a larger antenna array or a more sophisticated \ac{tas} transmission scheme is required in the densely overheard network. For instance, a relative secrecy cost around $0.15$ is achieved in the sparsely overheard network via both the schemes even with less than $M=2K = 32$ transmit antennas. To achieve such a level of secrecy cost, one requires to employ Scheme B with around $M= 64 K = 1024$ antennas. This observation explains the impact of \textit{slow} asymptotic decay. In fact under \ac{tas}, the system performance is still sensitive to the density of eavesdroppers, and hence more sophisticated \ac{tas} protocols and digital precoding schemes are required in networks with high density of malicious terminals.

\section{Secrecy For Free Under HADP}
\label{sec:hadp}
We now extend the analysis to \ac{mimo} architectures with \ac{hadp}. The main difference in this case is that the analog beamformers are implemented via \textit{phase shifters} instead of \textit{switches} and hence are of the form given in \eqref{eq:analog_HADP}. Similar to \ac{tas}, the analysis depicts that without using the \ac{csi} of the channels to eavesdroppers, secrecy can be achieved asymptotically for free. The relative secrecy cost vanishes significantly faster in this case. This is a direct result of employing all the antenna elements in \ac{hadp}. The derivations interestingly show that this behavior is obtained even by \textit{pure analog beamforming}. This agrees with intuitions given earlier in Section~\ref{sec:SFF}.

\subsection{Main Results}
Proposition~\ref{theo:2} gives a set of \textit{sufficient} conditions for achieving secrecy for free via \ac{hadp}, as well as a lower bound on the asymptotic decay of secrecy cost:
\begin{theorem}[Secrecy for free under HADP]
\label{theo:2}
Consider an \ac{hadp} architecture with $L \geq K$ \ac{rf} chains in which the \ac{bs} uses only the \ac{csi} of the legitimate channels for analog~and~digital~beamforming. Let analog beamformers be updated once per coherence time interval. Then, there exist linear digital~beamformers which achieve secrecy asymptotically for free. The asymptotic decay of secrecy cost in this case is at least $\log M$.
\end{theorem}

Similar to Proposition~\ref{theo:1}, this result describes sufficiency and does not discuss any necessary condition. It further implies that with \textit{the same number of \ac{rf} chains}, \ac{hadp} can tend the relative secrecy cost to zero significantly faster than \ac{tas}. In the sequel, we give a proof for this proposition by designing analog and digital beamformers which result in a relative secrecy cost decaying proportional to $\log M$. The derivations interestingly show that this property is achieved by only performing \textit{analog beamforming}. As \textit{fully digital precoding} is a special case of \ac{hadp}, Proposition~\ref{theo:2} further  guarantees the asymptotic secrecy for free in fully digital massive \ac{mimo} settings with the decay of order $\log M$. This agrees with the earlier derivations \cite{schaefer2017physical,bereyhi2018robustness}.

\subsection{Derivations}
The derivations follow the same approach as the one taken to prove Proposition~\ref{theo:1}. We start the proof by setting analog and digital beamformers. In the proposed scheme, we consider architectures with exactly $L=K$ \ac{rf} chains. The result for scenarios with $L>K$ \ac{rf} chains is then directly concluded, since the proposed scheme can also be used in those scenarios.

\begin{scheme}
Consider a hybrid architecture with exactly $L=K$ \ac{rf} chains. In the base-band domain, the \ac{bs} employs linear beamformers %
$\bw_k = \be_k^K$. %
The digital base-band signal is then coupled in the \ac{rf} domain using analog beamformers $\bff_1,\ldots,\bff_K$ whose entries for $k\in\dbc{K}$ and $m\in\dbc{M}$ read
\begin{align*}
\mathrm{f}_{k,m} = \frac{1}{\sqrt{M}} \frac{h_{k,m}^*}{ \abs{h_{k,m}} }.
\end{align*}
\end{scheme}

In this scheme, the digital beamformers simply map the information symbol of the~$k$-th~legitimate terminal to the $k$-th entry of the base-band signal. The analog beamforming then linearly combines the information symbols of different terminals using the corresponding channel phases. In other words, the beamforming in this scheme is performed \textit{purely via the analog beamformers}.

Considering the \ac{hadp} scheme, we have 
\begin{align}
\bh_k^\trp \mF \bw_i = \bh_k \bff_i = \frac{1}{\sqrt{M}} \sum_{m=1}^M \frac{h_{k,m} h_{i,m}^*}{ \abs{h_{i,m} } }
\end{align}
which for $i=k$ reduces to
\begin{align}
\bh_k^\trp \mF \bw_k = \frac{1}{\sqrt{M}} \sum_{m=1}^M { \abs{h_{i,m} } } = \frac{ \norm{\bh_k}_1 }{\sqrt{M}}.
\end{align}
As a result, $\sinr_k\brc{M}$ for $k\in \dbc{K}$ reads
\begin{align}
\sinr_k \brc{M} &= \dfrac{1}{M} \dfrac{ \displaystyle P_k \beta_k \norm{ \bh_{k} }_1^2  }{ \sigma^2 + \beta_k \tilde{I}_k }
\label{eq:sinr_hd}
\end{align}
with $\tilde{I}_k$ being defined as
\begin{align}
\tilde{I}_k = \dfrac{1}{M} \sum_{i=1 , i\neq k}^K P_i \abs {\sum_{m=1}^M \frac{h_{k,m} h_{i,m}^*}{ \abs{h_{i,m} } }}^2.
\end{align}
The leakage to the eavesdroppers is further given by $\esnr_k \brc{M} = {P_k \tilde{E}_k }/{\rho^2}$, where
\begin{align}
\tilde{E}_k = \dfrac{1}{M} \sum_{j=1}^J \theta_j \abs {\sum_{m=1}^M \frac{g_{j,m} h_{k,m}^*}{ \abs{h_{k,m} } }}^2.
\end{align}

To determine the asymptotic expansion of $\sinr_k\brc{M}$ and $\esnr_k\brc{M}$, we note the following statements:
\begin{itemize}
\item[(a)] $\abs{h_{k,m}}$ for $m\in \dbc{M}$ are \ac{iid} Rayleigh distributed random variables. Following the \ac{lln}, we have
\begin{align}
\dfrac{1}{M} \norm{ \bh_{k} }_1 \longrightarrow \Ex{\abs{h_{k,m}}}{} = \sqrt{\frac{\pi}{4} }.
\end{align}
\item[(b)] For $k\neq i \in \dbc{K}$, let the random sequence $\set{X_{k,i}\brc{m}}$ be defined as
\begin{align}
X_{k,i}\brc{m} = \frac{h_{k,m} h_{i,m}^*}{ \abs{h_{i,m} } }.
\end{align}
Noting that $\bh_k$ and $\bh_i$ are \ac{iid} vectors, it is concluded that $\set{X_{k,i}\brc{m}}$ is an \ac{iid} sequence whose entries have the following properties:
\begin{enumerate}
\item The expected value of $X_{k,i}\brc{m}$ reads
\begin{align}
\Ex{X_{k,i}\brc{m}}{} = \Ex{ h_{k,m} }{} \Ex{\frac{ h_{i,m}^*}{ \abs{h_{i,m} } }}{} = 0
\end{align}
\item The variance is given by
\begin{align}
\Ex{ \abs{X_{k,i}\brc{m}}^2 }{  } = \Ex{ 
\abs{h_{k,m}}^2 
 }{} = 1
\end{align}
\item For $i\neq j$, the covariance between $X_{k,i}\brc{m}$ and $X_{k,j}\brc{m}$ reads
\begin{align}
\Ex{ X_{k,i}\brc{m} X_{k,j}\brc{m}^* }{  } &= \Ex{ 
\frac{ \abs{h_{k,m}}^2 h_{i,m}^* h_{j,m} }{ \abs{h_{i,m} } \abs{h_{j,m} } }
 }{}\\
 &=
 \Ex{ \abs{  h_{k,m} }^2  }{} 
 \Ex{ \frac{  h_{i,m}^* }{ \abs{h_{i,m} }  }  }{} 
 \Ex{ \frac{  h_{j,m} }{ \abs{h_{j,m} }  }  }{} 
 \stackrel{\dagger}{=} 0
\end{align}
where $\dagger$ follows the fact that ${  h_{i,m}^* }/{ \abs{h_{i,m} }  }$ is a uniform random variable on the unit circle.
\end{enumerate}

Using the \textit{central limit theorem}, it is hence concluded that random variable $S_{k,i}$, defined as
\begin{align}
S_{k,i} \coloneqq  \frac{1}{\sqrt{M}} \sum_{m=1}^M X_{k,i}\brc{m} 
\end{align}
for $k\neq i\in\dbc{K}$, converges in distribution to a zero-mean and unit-variance~Gaussian~random variable. Following the fact that $X_{k,i}\brc{m}$ and $X_{k,j}\brc{m}$ are uncorrelated random variables, it is further straightforward to show that %
$\Ex{ S_{k,i} S_{k,j}^* }{  }  =  0$. %
for $i \neq j$. This means that random variables $S_{k,i}$ for $i \neq k \in \dbc{K}$ are \textit{asymptotically independent} and \textit{Gaussian distributed}.
\item[(c)] With the exact lines of derivations as in Statement (b), it is concluded that for $j \in \dbc{J}$
\begin{align}
Y_{k,j} \coloneqq \frac{1}{ \sqrt{M} } \sum_{m=1}^M \frac{g_{j,m} h_{k,m}^*}{ \abs{h_{k,m} } }
\end{align}
converge in distribution \textit{independent Gaussian variables} with zero mean and unit variance.
\end{itemize}

From Statement (b), we can write
\begin{align}
\tilde{I}_k = \sum_{i=1 , i\neq k}^K P_i \abs { S_{k,i} }^2
\end{align}
whose mean reads
\begin{align}
\Ex{ \tilde{I}_k}{} &= \sum_{i=1 , i\neq k}^K P_i  = P-P_k.
\end{align}
Its variance is further given by
\begin{align}
\Var{\tilde{I}_k} &= \sum_{i=1 , i\neq k}^K P_i^2 + \brc{\sum_{i=1 , i\neq k}^K P_i}^2 - \abs{\Ex{ \tilde{I}_k}{}}^2 = \sum_{i=1 , i\neq k}^K P_i^2 \leq P^2.
\end{align}
This indicates that $\tilde{I}_k$, for any $M$, is a random variable with bounded mean and variance. Similarly by using Statement (c), we have
\begin{align}
\tilde{E}_k = \sum_{j=1}^J \theta_j \abs { Y_{k,j} }^2.
\end{align}
The mean and variance of $\tilde{E}_k$ are hence straightforwardly calculated as
\begin{subequations}
\begin{align}
\Ex{ \tilde{E}_k}{} &= J \Ex{\theta}{ \rmF\brc{\theta} }\\
\Var{ \tilde{E}_k} &= J\Ex{\theta^2}{\rmF\brc{\theta}}
\end{align}
\end{subequations}
which are \textit{bounded} for any $M$.

We now utilize Statement (a) and write
\begin{align}
\left. \frac{1}{M} \right. \sinr_k \brc{M} &= \dfrac{ P_k \beta_k  }{ \sigma^2 + \beta_k \tilde{I}_k } \brc{\dfrac{\norm{ \bh_{k} }_1}{M}}^2 \longrightarrow \dfrac{ P_k \beta_k  }{ \sigma^2 + \beta_k \tilde{I}_k } \frac{\pi}{4}
\end{align}
for some realization of $\tilde{I}_k$. Noting that $\tilde{I}_k$ has bounded mean and variance, we can conclude that as $M$ grows large
\begin{align}
\sinr_k \brc{M} &= \tilde{\Gamma}_k M + \tilde{\epsilon}_k \brc{M}
\end{align}
where $\tilde{\Gamma}_k$ is a bounded constant with probability one, and $\tilde{\epsilon}_k \brc{M} = \setO\brc{1}$. For $\esnr_k\brc{M}$, we can further use the fact that $\tilde{E}_k$ has bounded mean and variance and write
\begin{align}
\esnr_k \brc{M} = \breve{\epsilon}_k\brc{m}
\end{align}
for large $M$, where $\breve{\epsilon}_k\brc{m} = \setO\brc{1}$ with probability one.

Using the asymptotic expansion of $\sinr_k\brc{M}$ and $\esnr_k\brc{M}$, it is concluded that for any realization of the channels, there exist some $M_0$, such that for $M>M_0$, the secrecy rate is non-zero. As a result, the relative secrecy cost reads
\begin{align}
\mathcal{C}\brc{M} \longrightarrow  \frac{ \displaystyle \sum_{k=1}^K q_k  \log \brc{ 
\breve{\epsilon}_k\brc{m}
} 
}{ \displaystyle \sum_{k=1}^K q_k \log\brc{ \tilde{\Gamma}_k M + \tilde{\epsilon}_k \brc{M}}} .
\end{align}
This concludes Proposition~\ref{theo:2}.


\subsection{Numerical Investigations}

In this section, we provide some numerical simulations to confirm our earlier derivations. In this respect, we consider the \textit{sparsely} and \textit{densely} overheard networks described in Section~\ref{eq:NumInv_TAS}. These networks are equipped with $K=16$ legitimate users, where in the former $J=2$, and in the latter $J=16$. We assume that the hybrid architecture contains $L=K$ \ac{rf} chains. The transmit power, noise variance, and channel coefficients are set similar to Section~\ref{eq:NumInv_TAS}, such that $\log \snr^\mm_k = 0$ dB and $\log \snr^\ee_j = -10$ dB for $k\in\dbc{K}$ and $j\in\dbc{J}$. 

In the sequel, we investigate two \ac{hadp} schemes in each of the considered networks:
\begin{itemize}
\item First, we give numerical results for the basic \ac{hadp} transmission scheme used for derivation of Proposition~\ref{theo:2}. We refer to this scheme as \textit{Scheme A}.
\item The investigations are then extended to a \textit{benchmark \ac{hadp} scheme}. 
To this end,~we~consider the low-complexity \ac{hadp} scheme proposed in \cite{liang2014low}. In this scheme, the analog beamformers are set proportional to the \textit{quantized} phase of the channel vectors, i.e., $\mathrm{f}_{k,m} = \exp\set{-\rmj \phi_{k,m}}/\sqrt{M}$ for $k\in\dbc{K}$ and $m\in\dbc{M}$, where $\phi_{k,m}$ is the phase of $h_{k,m}$ quantized via the least-squares method using $B$ bits. The digital beamformers are then given via \ac{zf} precoding over the effective channels, i.e., $\mF^\trp \mH$. More detailed discussions on this \ac{hadp} scheme can be followed in \cite{liang2014low}. This scheme is referred to as \textit{Scheme B}.
\end{itemize}
For sake of simplicity, we assume uniform power allocation in these schemes and set $q_1 = \ldots = q_K = 1$ throughout the simulations.

\begin{figure}
\centering
\input{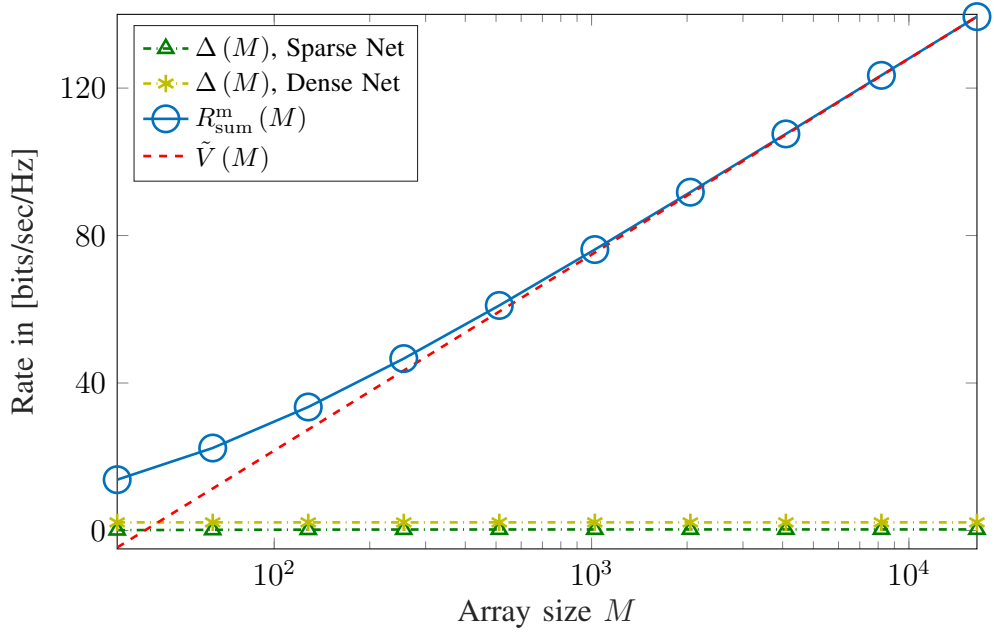}
\caption{Achievable sum-rate and information leakage for \ac{hadp} Scheme A.}
\label{fig:5}
\end{figure}

Fig.~\ref{fig:5} shows the achievable sum-rate, as well as the information leakage in the network, against the array size $M$ for \ac{hadp} Scheme A. As the figure shows, the information leakage remains constant in $M$. Comparing the results with the counterpart \ac{tas} architecture, one observes a significant performance gain in this case. This follows the fact that the achievable rate to each legitimate terminal under \ac{hadp} grows proportional to $\log M$ which is significantly faster than the growth speed under \ac{tas}, i.e., $\log \log M$. As a result, the approximation $R_\avg^\mm\brc{M} \approx R_\avg\brc{M}$ is accurate at smaller values of $M$ in this case. The secrecy cost decaying speed implies that 
\begin{align}
R_\avg^\mm \brc{M} \approx \tilde{V} \brc{M} = \tilde{R}_0 + \tilde{T} K \log M 
\end{align}
for some constants $\tilde{R}_0$ and $\tilde{T}$. For the result given in Fig.~\ref{fig:5} this asymptotic curve is numerically fitted. As it is seen, the fitted curve tracks closely the simulations.
\begin{figure}
\centering
\input{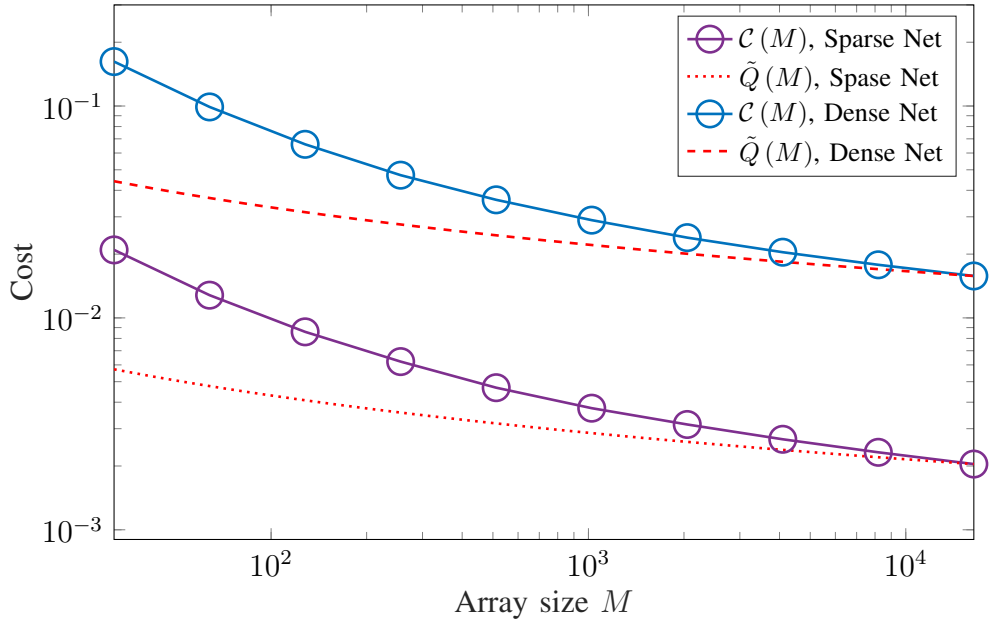}
\caption{Relative secrecy cost for \ac{hadp} Scheme A.}
\label{fig:8}
\end{figure}


The resulting cost function for the scenario in Fig.~\ref{fig:5} is plotted in Fig.~\ref{fig:8}. Similar to the previous figure, we use the asymptotic expansion and write the large-system approximation
\begin{align}
\maC \brc{M} \approx \tilde{Q} \brc{M} = \frac{\tilde{\epsilon}_0}{\log M}
\end{align}
for some constant $\tilde{\epsilon}_0$. The fitted curves are further shown in the figure. As it is observed, the curves give better approximations at large values of $M$. 

Similar to the scenarios with \ac{tas}, the increase in the density of malicious terminals in the network results in higher information leakage. Comparing this degradation in secrecy performance under \ac{tas} and \ac{hadp}, the latter shows less sensitivity to the density of the malicious terminals in the network. This is, as in \ac{hadp} all the transmit antennas are active. As a result, the analog beamformers can construct narrower downlink beams compared to the transmitters which use purely digital beamforming with \ac{tas}.

\begin{figure}
\centering
\input{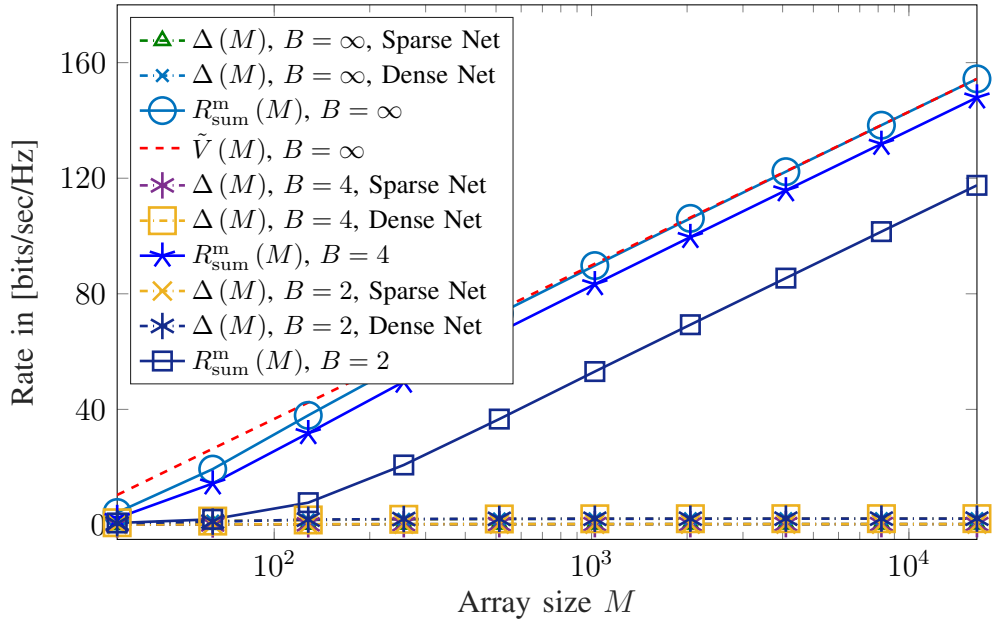}
\caption{Achievable sum-rate and information leakage for \ac{hadp} Scheme B.}
\label{fig:10}
\end{figure}

\begin{figure}
\centering
\input{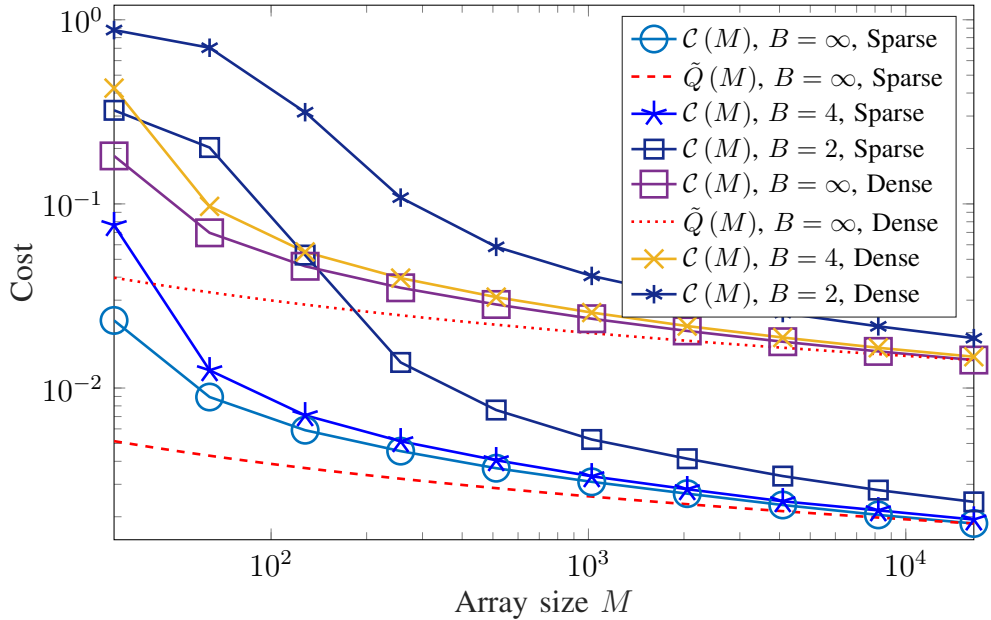}
\caption{Relative secrecy cost in the sparsely overheard network for \ac{hadp} Scheme B.}
\label{fig:9}
\end{figure}

The simulation results for Scheme B are further given in Figs.~\ref{fig:10} and \ref{fig:9} considering various numbers of bits for quantization of phase-shifts in analog beamformers. In Fig.~\ref{fig:10}, the achievable sum-rate and the information leakage are plotted against the array size. The figure shows, the achievable rate, in this case for the benchmark scheme with infinite quantization accuracy $B\uparrow \infty$, is improved compared to the basic scheme. This is due to the fact that Scheme B performs digital precoding in addition to analog beamforming. $\Delta\brc{M}$ however remains unchanged in this case. 

Fig.~\ref{fig:10} further shows the performance for more practical scenarios with finite quantization accuracy, namely $B=2$ and $B=4$. From the figure one can see that low-resolution quantization only shifts the achievable sum-rate and does not change the asymptotic decay of secrecy cost. For the information leakage, quantization does not result in any sensible change which is intuitive. In fact, quantization only introduces performance degradation. Since with high-resolution quantizers, the information leakage is close to zero, further degradation is not negligible in this case.

The higher achievable sum-rate in Scheme B results in lower secrecy costs. This is shown in Fig.~\ref{fig:9} where we sketch the relative secrecy cost against the array size for the sparsely and densely overheard networks, respectively. Comparing the variations in the two figures against the benchmark scheme under the \ac{tas} approach, one observes that the system in this case is less sensitive to the density of eavesdroppers. 


\section{Conclusions}
\label{sec:Conc}
Secrecy is achieved at almost no cost in \ac{mimome} settings with large transmit arrays. We call this property \textit{secrecy for free} and find that it is generic. This means that regardless of the transceiver architecture, the achievable downlink rate to the legitimate users remains almost the same in the presence or absence of malicious terminals, when the number of transmit antennas is large enough. Despite the generality of this property, the decaying speed of the \textit{information leakage per achievable rate} varies from one architecture to another. Our investigations have demonstrated that using a conventional \ac{hadp} scheme, this ratio converges to zero significantly faster than \ac{mimome} settings which utilize \ac{tas} for \ac{rf} cost reduction. Noting that \ac{hadp} schemes with low-resolution quantized phase shifters are more trivial to implement, it is concluded that \ac{hadp} is an effective hybrid architecture for massive \ac{mimo} implementation in practice. 


The derivations in this work have been given for downlink transmission schemes which do not use the \ac{csi} of overhearing channels. The results hence imply that in massive \ac{mimome} settings, physical layer security is achieved even without taking the malicious terminals into account for system design. In other words, even by using conventional channel coding techniques, eavesdroppers can be effectively suppressed via \textit{good} downlink beamformers. This agrees with the intuition from the \textit{favorable propagation property} which indicates that proper beamforming via large transmit arrays can almost blind all other receiving terminals in the network.

The current work can be extended in various respects. A natural direction is to study available \ac{hadp} designs to find out the most effective approach which achieves rather small secrecy cost at practical dimensions. To this end, one needs to extend the derivations to other channel models, e.g. the models proposed for \ac{mmW} communications. Although such studies seem to be straightforward extensions of this work, they give more insights on the secrecy performance of massive \ac{mimo} systems with practical dimensions.

\begin{acronym}
\acro{mimo}[MIMO]{multiple-input multiple-output}
\acro{hadp}[HADP]{hybrid analog-digital precoding}
\acro{mimome}[MIMOME]{multiple-input multiple-output multiple-eavesdropper}
\acro{csi}[CSI]{channel state information}
\acro{lln}[LLN]{law of large numbers}
\acro{awgn}[AWGN]{additive white Gaussian noise}
\acro{iid}[i.i.d.]{independent and identically distributed}
\acro{ut}[UT]{user terminal}
\acro{bs}[BS]{base station}
\acro{mt}[MT]{mobile terminal}
\acro{tas}[TAS]{transmit antenna selection}
\acro{lse}[LSE]{least squared error}
\acro{rhs}[r.h.s.]{right hand side}
\acro{lhs}[l.h.s.]{left hand side}
\acro{wrt}[w.r.t.]{with respect to}
\acro{tdd}[TDD]{time-division duplexing}
\acro{mrt}[MRT]{maximum ratio transmission}
\acro{zf}[ZF]{zero forcing}
\acro{rzf}[RZF]{regularized zero forcing}
\acro{snr}[SNR]{signal-to-noise ratio}
\acro{sinr}[SINR]{signal-to-interference-and-noise ratio}
\acro{rf}[RF]{radio frequency}
\acro{mf}[MF]{match filtering}
\acro{mmse}[MMSE]{minimum mean squared error}
\acro{qos}[QoS]{quality of services}
\acro{mmW}[mmWave]{millimeter Wave}
\end{acronym}

\bibliographystyle{IEEEtran}
\bibliography{ref}
\end{document}